\documentclass{aa}             
\usepackage{amssymb,graphicx}  
       
\begin{document}
\title {Early humps in WZ Sge stars}
\author{Yoji Osaki \inst{1}
\and Friedrich Meyer \inst{2}}
\offprints{Yoji Osaki; osaki@net.nagasaki-u.ac.jp}

\institute{Faculty of Education, Nagasaki University, Nagasaki
 852-8521, Japan 
\and
 Max-Planck-Institut f\"ur Astrophysik, Karl Schwarzschild Str. 1, 
D-85740 Garching, Germany}

\date{Received: / Accepted:}

\abstract{
Photometric humps in outburst that are locked with the binary orbital
period have been
observed exclusively in the early phase of outbursts of WZ Sge stars. 
It is suggested that this "early hump" phenomenon is the 
manifestation of the tidal 2:1 resonance in accretion disks of
binary systems with extremely low mass ratios. The "early humps" can 
be understood by the two-armed spiral pattern of tidal dissipation
generated by the 2:1 resonance, first discussed by Lin \& Papaloizou
(1979). The tidal removal of angular momentum from the disk during
outbursts of dwarf novae, an important feature, is discussed in the
context of the disk instability model. The 
ordering of tidal truncation radius, the 3:1 and 2:1 resonance
radius in systems of different mass ratio naturally leads to a
classification of dwarf nova systems in three groups according
to their mass ratio. The WZ Sge stars are those systems which
have the lowest mass ratios and are therefore characterized by "early humps".
\keywords{accretion disks -- cataclysmic variables  -- stars: dwarf novae
--  stars individual: WZ Sge}
}
\authorrunning{Yoji Osaki \& Friedrich Meyer}
\titlerunning {Early humps in WZ Sge stars}
\maketitle

\section{Introduction}
Several different types of periodic photometric humps are observed in
light curves of dwarf novae both in outbursts and in quiescence. The
most well known among them are the "orbital humps" in 
quiescence, which are repeated with the binary orbital period, and
the "superhumps" observed during the superoutburst of SU UMa-type
dwarf novae, which are repeated with a period slightly longer than 
the binary orbital period. Both types of humps are now well
understood; the orbital hump is due to the
phase-dependent visibility of the hot spot located at that 
place of the accretion disk where the in-falling stream from the
secondary star collides with the disk rim while the superhump is due
to intrinsic light variation produced by 
tidal dissipation on an eccentric precessing disk in SU UMa stars
(see e.g. Warner 1995). 

Different from these is another type of photometric humps recently
recognized as "early humps"
observed in the first about ten days of the outbursts in WZ Sge
stars. These humps repeat with the binary orbital period but 
they only appear in the early phase of outburst in WZ Sge-type dwarf
novae exclusively and they are replaced by the ordinary
"superhumps" in the later phase of the outburst. The "early 
hump" phenomenon was first discovered in the 1978 outburst of WZ Sge
itself by Patterson et al. (1981) and has been observed
consequently in
other WZ Sge systems: the 1995 outburst of AL Com (Kato et 
al. 1996; Patterson et al. 1996); the 1992 outburst of HV Vir
(Leibowitz et al. 1994) and the 1996 outburst of EG Cnc (Matsumoto 
1998).
These "early humps" have either been called "outburst orbital
humps" by Patterson et al. (1996) or "early superhumps" by
Kato et al. (1996). The different designation reflects a
different interpretation of the phenomenon: Patterson et al. (1981, 
1996)
favor a "super-hot spot" interpretation in which early humps are due to
a brightened hot spot which in turn is due to an
enhanced mass transfer from the secondary star during outburst 
(Patterson et al. 1981),  while Kato et al. (1996) favor an
"early superhump" interpretation in which early humps are a premature
form of the true superhumps. Here we call these humps simply  
"early humps" to avoid any particular interpretation. 

In July 2001 WZ Sge (the proto-type of its group) underwent an 
unexpected
full-scale outburst, 10 years earlier than expected
with its former 33 years recurrence period. The star was caught on
the rising branch of its outburst light curve. It was extensively
observed by professional and amateur astronomers. Beautiful light curves
showing "early humps"  with an initial amplitude reaching 0.5 mag
and then decreasing a few tenths were observed
(VSNET 2001:http://www.kyoto-u.ac.jp/vsnet/DNe/wzsge01.html).  

In this investigation we first suggest that the early humps exclusively
observed in the early phase of outbursts in WZ Sge stars are most
likely a manifestation of the 2:1 resonance in the accretion disk
in these systems with extremely low mass ratios 
(Sect.2). The amplitude of the early humps in the light curve is
discussed in Sect. 3. In Sect. 4 we then present a general
discussion of angular momentum removal from the accretion disk during
outbursts of dwarf novae based on the disk instability model.
This leads (Sect. 5) to a new subdivision in
the dwarf novae unification model
of Osaki (1996) which allows to distinguish between ordinary SU UMa 
stars and
WZ Sge stars as those with the lowest mass ratios exhibiting
"early humps" and "echo outbursts".   

\section {Early humps: a  manifestation of the    
tidal 2:1 resonance}

One of peculiarities of WZ Sge stars is the rather late appearance of 
the
superhumps. In the 1987 outburst of WZ Sge, the regular superhumps
first appeared ten days after the start of the outburst, while
in ordinary SU UMa stars they usually develop within only a few days. 
The late 
development of superhumps in WZ Sge stars can be understood by the low
growth rate of the eccentric tidal instability in these systems. Lubow
(1991) showed analytically that the eccentricity growth rate is
proportional to the square of the binary mass ratio $q$, if other
conditions are kept the same. This very slow growth in the case
of extremely low mass ratio systems is also confirmed by numerical
simulations (Hirose \& Osaki 1990; Whitehurst 1994).  

If in WZ Sge stars the eccentric disk is not yet well developed in
the early phase when the major outburst has already started, the disk 
must
expand well beyond the 3:1 resonance radius because nothing 
removes the angular momentum (released by the accreting matter).
As demonstrated below, in binary systems with the extremely low 
mass ratio 
of WZ Sge stars, the disk would expand well beyond another resonance radius, 
the 2:1 resonance radius. In such a case the 2:1 resonance acts
to truncate the disk. The 2:1 resonance is a very strong resonance as 
the 
periodic tidal force acting upon the disk resonates with the two-armed 
(m=2) wave pattern in the Keplerian disk, i.e., the inner Lindblad 
resonance. 

In fact, more than
twenty years ago, Lin \& Papaloizou (1979) showed that the
accretion disk in binary systems with extremely low mass ratio is
truncated at the 2:1 resonance radius and that near the resonance,
a strong two-armed dissipation pattern forms.
We here suggest that the tidal effect due to the 2:1 resonance is
responsible for the "early humps" observed in WZ Sge stars.

The two-armed dissipation pattern produced by the 2:1 resonance 
was shown in Fig. 3a in Lin \& Papaloizou (1979). From 
this figure we can find that the strongest dissipation appears around  0.7 in 
the binary orbital phase while the second peak appears around 0.2 
where phase zero is defined at the conjuction of the secondary 
star in front of the primary white-dwarf, i.e., the eclipse center. 
This is in good agreement with the observed phases of the 
double humps in the 2001 
WZ Sge outburst. On the other hand, in the super-hot spot model 
proposed by Patterson et al. (1981), the spot in outburst would have to 
be displaced  by 60 degrees to the downstream direction from its 
position at quiescence, a rather unlikely possibility. The 
double-hump nature of the early humps also is difficult to explain 
by the hot spot model.  The interpretation of early humps as a 
premature form of superhumps is clearly ruled out because early 
humps are repeated with the binary orbital period and not with 
the ordinary superhump period which is always longer than the 
orbital period by about one percent (or a few percent in the case 
of ordinary SU UMa stars) and because the amplitudes of the early 
humps were found to be larger than that of the ordinary superhumps 
in the 2001 WZ Sge outburst. Furthermore, as to the two other 
interpretations of the early humps, i.e., the super-hot spot model 
and the early superhump model, it has remained to be demonstrated 
why the early humps appear exclusively in those systems with extreme 
mass ratios.

\section {The amplitude of the early humps in the light curve}

In the 2001 outburst of WZ Sge, large-amplitude periodic humps were 
observed 
near the maximum with an initial amplitude of 0.5 mag but then
settling to an amplitude of a few tenths of a magnitude. Basically
the same phenomenon was observed in the 1995 outburst of AL Com 
but 
with a much lower amplitude.  
We interpret this phenomenon as a two-step process, an initial 
transient 
adjustment stage of the disk and a more or less quasi-steady state
in the later stage.   

We first discuss the initial transient stage. Let us consider 
what would happen if the quiescent disk suddenly turns to a fully viscous 
state. 
In our picture of WZ Sge stars, the viscosity in the quiescent disk is 
extremely 
low. In the extreme case, material transfered from the secondary star is 
accumulated simply in a torus with its radius given by the 
circularization 
radius (Lubow-Shu radius). The circularization radius is 
given by 
\begin{equation}
R_{\rm cir}/a=(1+q)(R_{\rm L_1}/a)^4,
\end{equation}
where $a$, $q=M_2/M_1$, $R_{\rm L_1}$ are the binary separation, the 
binary 
mass ratio, and  the distance from the primary white dwarf to the 
inner Lagrangian point, respectively. The distance to the Lagragian 
point, 
$R_{\rm L_1}$, is given by (see, e.g., Warner 1995) 
\begin{equation}
R_{\rm L_1}/a=1-w+\frac{1}{3}w^2+\frac{1}{9}w^3
\end{equation}
where 
\begin{equation}
w^3=\frac{q}{3(1+q)}\quad  {\rm for} \quad q\leq 0.1.  
\end{equation}

When an outburst occurs, the torus spreads out into the disk. 
Under the assumption of angular momentum conservation the outer edge
of the resultant disk will then be given  
by $R_{\rm outer~edge}/a=(7/5)^2 R_{\rm cir}/a$.
Here we have assumed the surface density profile $\Sigma=r^{-3/4}$ of 
a steady 
hot disk with free-free opacity whose mean specific angular momentum, 
$j_{\rm m}$, is given by $j_{\rm m}=(5/7)\sqrt{GM_1 R_{\rm 
outer~edge}}$.
This is to be compared with the 2:1 resonance radius given by
$R_{\rm 2:1}/a=(2^2(1+q))^{-1/3}.$

Figure 1 illustrates the outer edge of the disk  
and the 2:1 resonance radius as a function of mass ratio. 
We find that the outer edge of the disk exceeds the 2:1 resonance radius 
in the low mass-ratio systems with $q$ less than 0.08. 
In those systems, the 2:1 resonance ensues, resulting in the two-armed 
spiral shocks and strong tidal torques acting on the disk. 
The extra angular momentum of the disk will be rapidly removed from 
the disk. We find from Fig.1 that the extra angular momentum to be
gotten rid off becomes greater as the binary mass ratio becomes smaller.
The tidal dissipation luminosity resulting from
the transfer of angular momentum by the 2:1 resonance is given by 
\begin{equation}
L_{\rm{tid}}= \left(\Omega_{\rm K}-\Omega_{\rm{orb}}\right)\dot I
\end{equation}
with $\Omega_{\rm K}=(GM_1/r_{\rm d}^3)^{1/2}$ Kepler angular
frequency, $\Omega_{\rm{orb}}=(GM/a^3)^{1/2}$ orbital angular
frequency, $M=(1+q)M_1$ total mass and $r_{\rm d}$ the outer disk 
radius, 
$ \dot I$ rate of angular momentum removed tidally from the accretion 
disk. At the earliest phase of the outburst, the rate of tidal 
removal of angular 
momentum from the disk, $ \dot I$, must be much larger than that of the 
later 
phase discussed below because of the extra angular momentum of the disk 
to be removed. 

After this intial transient phase is ended, a slower removal of 
angular 
momentum by the 2:1 resonance will follow. Let us estimate 
the amplitude of the early hump in this second stage
expected from this 2:1 resonance model. 

\begin{figure}[ht]
\begin {center}
\includegraphics[width=6.0cm]{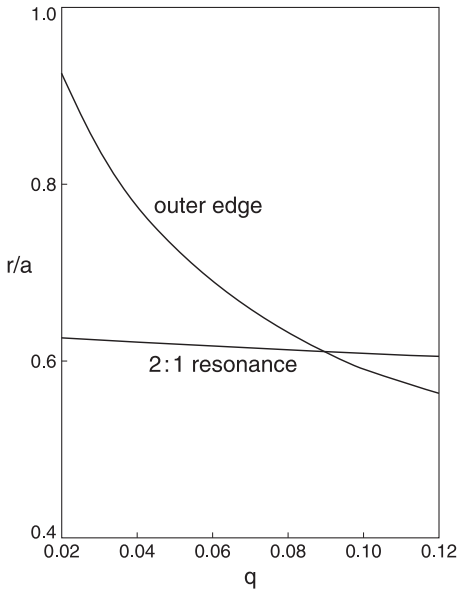}
\caption {Outer edge of the disk and radius of the 2:1
resonance, measured in units of binary separation $a$, as a function
of the mass ratio $q$.
}
\end {center}
\end{figure}       

For this we compare the 2:1 resonance luminosity (i.e. the
non-axisymmetric component of the dissipation) with the luminosity
of the accretion disk (i.e. the axisymmetric component of the
dissipation)
\begin{equation}
L_{\rm{d}}=\frac{1}{2} \frac{GM_1 \dot M}{r_{\rm {WD}}}
\end{equation}
where $r_{\rm {WD}}$ is the radius of the central white dwarf. In
steady state the rate
$\dot I$ of angular momentum transfer is equal to the rate of angular
momentum released in the disk,
\begin{equation}
\dot I=\dot M \sqrt{GM_1 r_{\rm m}}.
\end{equation}
Here $r_{\rm m}$ is that radius where the Kepler specific angular
momentum is equal to mean specific angular momentum of the disk. For
the surface density profile $\Sigma \sim r^{-3/4}$ as before one
obtains $r_{\rm m}={(\frac{5}{7})}^2 r_{\rm d}$. At the 2:1 resonance
$\Omega_{\rm {orb}}$ is $\Omega_{\rm K}/2$. Inserting these values
into Eq. (4) we obtain 
\begin{equation}
L_{\rm{res}}=\frac{5}{14} \frac{GM_1 \dot M}{r_{\rm d}}.
\end{equation}
We then find that the ratio of the 2:1 resonance luminosity to the
disk luminosity is
\begin{equation}
\frac{L_{\rm{res}}}{L_{\rm d}} = \frac{5}{7} \frac {r_{\rm {WD}}}
{r_{\rm d}} \simeq 0.03
\end{equation}
where we took $r_{\rm {WD}}= 10^9$cm, $r_{\rm d}=
10^{10.4}$cm.
This value is already of the right order of magnitude to explain the
amplitude of the early humps observed in other WZ Sge systems, e.g.
AL Com : 0.05 mag, HV Vir : 0.1 mag, EC Cnc: 0.02 mag, provided that
the bolometric corrections for the disk luminosity and for the resonance
luminosity are similar.

The following consideration estimates the relations in more detail.
By integrating the bolometrically corrected (Allen 1973) contributions of
individual rings of a standard accretion disk and comparing the
resulting visual luminosity with the bolometric one one can derive a
bolometric correction for the disk as a whole. 

For the parameters as above and an estimated accretion rate of $\dot M
= 10^{-7} M_\odot$/yr for the early outburst phase of WZ Sge we obtain
\begin{equation}
\frac{({L_{\rm d})}_{\rm V}}{L_{\rm d}}=10^{-0.4{(\rm{BC})}_{\rm d}}
\end{equation}
with $(\rm{BC})_{\rm d} \simeq 3.8$ where the subscript V denotes the 
visual luminosity and BC is the bolometric correction.

For the 2:1 resonance luminosity the bolometric correction is 
determined by the effective temperature
which is obtained from the luminosity and the area from which it is
radiated. For the latter
we estimate a total azimuthal extent of $\pi r$ (see Lin \& Papaloizou
1979) and assume a width of 4$H$ where $H$ is the scaleheight for the
obtained temperature. With the parameters of WZ Sge as above this
gives $(T_{\rm{eff}})_{\rm{res}} \simeq 10^{4.2}$ K and a bolometric
correction $(\rm{BC})_{\rm {res}} \simeq 2.6$.

The ratio of the visual luminosity from the resonance dissipation to
that from the disk is then,
larger than that of the bolometric luminosities by the factor
$10^{-0.4[(\rm{BC})_{\rm{res}}-(\rm{BC})_{\rm d}]} \simeq 3$. With
Eq. (8) we obtain 
\begin{equation}
\frac{({L_{\rm {res}})}_{\rm V}}{({L_{\rm d})}_{\rm V}}=\simeq 0.09.
\end{equation}

In addition a geometrical effect has to be considered.
In order to determine the hump
amplitude we have to know the orientation of the radiating surfaces
with respect to the direction towards the observer. We assume
that half of the spiral dissipation pattern surface faces upwards
and downwards and half points radially. We account for a 
slightly uneven distribution of the two armed spiral dissipation
pattern by attributing 2/3 of the tidal dissipation to
the major arm and 1/3 to the minor arm. Maximum light will be seen at
the orbital phase when the major arm is facing in our direction,
minimum light about a quarter of the period later when the surfaces
pointing radially are invisible. This gives a relative amplitude of
the luminosity variation
\begin{equation}
\frac{\frac{2}{3}(L_{\rm{res}})_{\rm V} \cdot \sin i}
{[(L_{\rm d})_{\rm V}+\frac{1}{2} (L_{\rm{res}})_{\rm V}]
\cos i} = 0.05 \tan i
\end{equation}
where $i$ is the inclination angle of the disk, $\pi/2 -i$ that of the
radially pointing surface at maximum.
The expected hump amplitude then becomes 
\begin{equation}
\Delta m_{\rm V}= 2.5 \log (1+ 0.05 \tan i)
\end{equation}
and for an inclination angle for WZ Sge of $i=75^\circ$
(Smak 1993)
\begin{equation}
\Delta m_{\rm V} \simeq 0.2,
\end{equation}
is in good agreement with the hump amplitude observed for WZ Sge of 
${(\Delta m_{\rm V})}_{\rm {WZ}}$ = 0.2 to 0.15 after the first two 
days.

WZ Sge is the only eclipsing system among the known WZ Sge stars.
The other non-eclipsing systems must have lower inclination,
leading to lower values for the hump amplitude, e.g.
$\Delta m_{\rm V}$= 0.1 for $i=60^\circ$ and
$\Delta m_{\rm V}$= 0.03 for $i=30^\circ$ . This fits well to the 
observed
range of $\Delta m_{\rm V}$ in these other systems mentioned above.

\section {Angular momentum removal from the disk in dwarf novae}
The cyclic behavior of outburst/quiescence in dwarf novae is most
naturally understood by the disk instability model (Osaki 1974). In
this model, mass supplied from the secondary star to the disk 
is not accreted onto the central white dwarf but rather is mostly
stored within the disk during quiescence. The material accumulated
in the disk is then suddenly accreted onto the central white 
dwarf during outburst due to an intrinsic instability within the
disk. The intrinsic instability has been identified as the thermal
instability in the disk, which produces a jump in disk 
viscosity (Meyer \& Meyer-Hofmeister 1981; for the general review of
theoretical aspects of the disk instability model, see, e.g.,
Cannizzo 1993 and Osaki 1996). 

Thus the cyclic behavior of dwarf novae is
well understood as a process of interchanging
mass accumulation and drainage in the disk. Equally important here
is the cyclic variation of the total angular
momentum in the disk. Angular momentum also accumulates
during quiescence of dwarf novae. When an outburst occurs, the increase
of viscosity leads to accretion onto the central white dwarf. 
However, in order to conserve angular momentum some material must move
outward to radii of higher Kepler angular momentum
giving rise to an expansion of the disk. 
The angular momentum brought outward will finally be returned to the
orbit of the binary by tidal torques from the secondary star acting
in the disk. It is thus
understood that in a time-average over the outburst cycle a steady
state is established for the total angular momentum of the disk in a
same way as for the total mass of the disk. 

In ordinary U Gem-type dwarf novae, a quasi-periodic
outburst/quiescence cycle repeats in which the disk expands to the 
tidal truncation radius during an outburst. There extra angular 
momentum can be returned to the binary orbital motion via tidal
torques. The accretion disk is then truncated
at this radius. During quiescence, the disk radius gradually decreases
since the matter added to the disk from the secondary star has a
specific angular momentum lower than that at the disk edge.
This gives rise to a cyclic
variation of the disk radius. This picture fits very well to the
observed variation of the disk radius in dwarf novae (Anderson 1988; 
Ichikawa \& Osaki 1992).

Paczynski (1977) calculated the tidal truncation radius as that of
the last non-intersecting orbit of a test particle around the central 
star in
the binary potential. His results agree fairly well with an analytic
treatment of the tidal torques in the binary by Papaloizou \& Pringle
(1977) who showed that tidal torques increase strongly with an increase
of the radial coordinate in the disk.

For the angular momentum removal
from the disk a new aspect emerged with the discovery of the tidal
instability by Whitehurst (1988). In this
instability, an accretion disk is deformed to an eccentric elliptic
shape and then precesses progradely in the inertial coordinate system. 
When the eccentric disk is developed a very efficient removal of
angular momentum from the disk is supposed
to occur. The 3:1 resonance in
the accretion disk, responsible for the tidal instability, is
only possible in binary systems with a ratio $q=M_2/M_1$ of the
secondary to the primary mass less than about 
0.25. This is because only in these low
mass-ratio systems the tidal truncation radius is large enough
to accommodate the 3:1 resonance radius within its boundary.  

By combining the tidal instability with the thermal instability,
Osaki (1989) proposed the thermal-tidal instability model (called TTI
model) for the superoutburst cycle of SU UMa-type dwarf novae. In this
model, the short normal outbursts occur as long as the disk radius
is small and the 
mass accreted during the normal outbursts is less than that accumulated
during quiescent intervals. Mass and angular momentum of the disk
gradually build up. Thus during the sequence of normal outbursts
the disk radius gradually increases until a final normal outburst
drives the outer edge
of the disk beyond the 3:1 resonance radius, triggering the tidal
instability. Enhanced tidal removal of angular momentum in the eccentric
precessing disk then keeps the disk in a hot state longer than a 
normal outburst does. This explains the long duration of the 
superoutburst and
the superhump phenomenon. The typical supercycle
length of SU UMa stars is a few hundred days, i.e., less
than about a year.   

The WZ Sge stars are an extreme case of SU UMa stars with long
supercycles lasting decades. Besides a long recurrence time, WZ Sge 
stars
exhibit several other unique characteristics, which are discussed by 
various workers (for modeling see Osaki 1995). It is thought that
these systems have
mass transfer rates lower than the ordinary SU UMa stars, $10^{15}$ g/s
for WZ Sge stars versus $10^{16}$ g/s for the ordinary SU UMa stars. 
During quiescence the disk viscosity is very low (Smak 1993)
which can be understood as to be related to the fact that the
secondary stars are brown dwarfs without magnetic activity (Meyer
\& Meyer-Hofmeister 1999),
mass ratios $q=M_2/M_1$ less than 0.1, most likely as low as 0.03.
The ``echo outbursts'', a repetitive rebrightening after the main
outburst, beautifully established during the 2001 outburst, are also
related to the low viscosity (Osaki et al. 2001).

Recently Hellier (2001) made an interesting suggestion that some of
peculiarities of WZ Sge stars and some of ER UMa stars can be
understood within the TTI model by considering a possibility of 
rather weak tidal torques in the eccentric disk, in those with extreme
mass ratios. He suggested that tidal dissipation in these systems is
too weak to sustain the disk in the hot state long enough.  A
premature shut-down of the superoutburst could be responsible for
"echo outburst" observed after the end of the main outburst in EG Cnc.
Indeed Osaki et al. (2001) showed that the viscosity decrease
related to the disk cooling after the main outburst and magnetic field
decay causes the repetitive rebrightening of EG Cnc. 

As discussed in section 3, in binary systems like WZ Sge stars 
with the extremely low mass ratio, of $q\leq 0.08$, once an outburst 
occurs, 
the outer edge of the hot viscous disk exceeds the 2:1 resonance radius, 
thus exciting the strong two-armed dissipation pattern due to the 2:1 
resonance, which now can remove angular momentum from the disk. 
The WZ Sge stars are thus those systems which are characterized 
by "early humps". Since the 3:1 resonance radius is smaller the 2:1 
resonance 
radius, the tidal eccentric instability due to the 3:1 resonance 
still operates in those systems even though its growth rate is rather 
low. 
Eventually (after about ten days from the outburst maximum) the 
precessing  
eccentric pattern characterized by "ordinary superhumps" has grown to a 
sufficient amplitude. "Early humps" are now replaced 
by "ordinary superhumps" because the disk's outer edge does 
not need to reach the 2:1 resonance radius any more as the tidal 
eccentric 
pattern due to the 3:1 resonance can now remove angular momentum 
from the disk. Our interpretation is that what is occurring 
during outbursts of WZ Sge stars is just this process. 

\section {New classification of cataclysmic variables}
In Fig. 2 we show the tidal truncation radius and the 3:1 resonance
radius as a function of the mass ratio (see also Hellier (2001))
and in addition the 2:1 resonance radius. This third line in our 
diagram,
the 2:1 resonance radius, now allows a new classification of CVs. CV
systems can be divided into three groups according to
their mass ratio: (1) high mass ratio systems in which the tidal
truncation radius is smaller than the 3:1 resonance radius
correspond to the U Gem-type dwarf novae, (2) intermediate mass ratio
systems in which the 3:1 resonance radius is smaller than the tidal
truncation radius, but the tidal truncation radius is smaller than the
2:1 resonance radius, correspond to the ordinary SU UMa stars, (3)
lowest mass-ratio systems in which the 2:1 resonance radius is between 
the tidal truncation radius and the 3:1 resonance radius correspond
to WZ Sge stars. 

\begin{figure}[ht]
\begin {center}
\includegraphics[width=6.0cm]{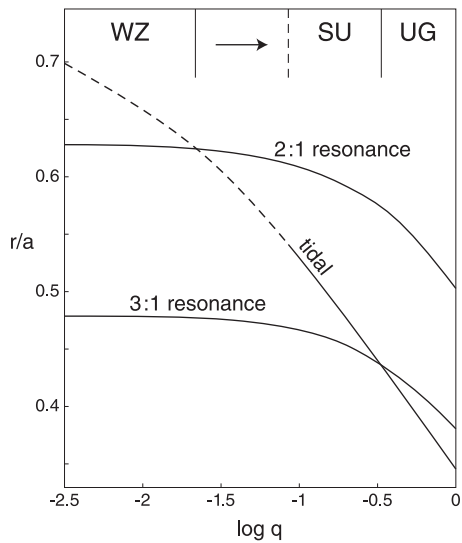}
\caption {Tidal truncation radius together with the radii of the 2:1
and 3:1 resonance, measured in units of binary separation $a$, as a function
of the mass ratio $q$ and dividing lines between the three classes of
systems. The dividing line between WZ Sge and SU Uma systems is moved
to the dashed position if the tidal force becomes to weak to truncate
the disk (see text).
}
\end {center}
\end{figure}

The exact location of the boundaries between the three cases is 
uncertain 
since the tidal truncation radius (taken here as 90\% of the radius of 
the Roche lobe) is rather approximate. 
In particular, the boundary between the SU UMa stars and the WZ Sge 
stars in 
Fig. 2 is found to be around $q=0.025$, which seems to be too small as 
compared with $q\simeq 0.08$ obtained in section 3. In those binary 
systems 
with the extreme low mass ratio, the tidal force will be weak and the 
ordinary tidal torques at the tidal truncation radius may not be strong 
enough to stop sudden expansion of the disk caused by an outburst. 
In such a case, the 2:1 resonance radius could be the only place where 
the disk is effectively truncated and the boundary between the SU UMa 
stars 
and the WZ Sge stars may be at $q\simeq 0.08$ rather than $q\simeq 
0.025$.

\begin{acknowledgements}
We would like to thank Emmi Meyer-Hofmeister for helpful discussions 
and 
technical assistance. 
Yoji Osaki acknowledges financial support from the Japanese Ministry of 
Education, Culture, Sports, Science and Technology with a Grant-in Aid 
for 
Scientific Research No. 12640237.

\end{acknowledgements}


\begin{thebibliography}
{}
\bibitem{ref:5}  Allen C.W., 1973, Astrophysical Quantities, 3rd
edition, University London, The Athlon Press
\bibitem{ref:10} Anderson N, 1988, ApJ 325, 266
\bibitem{ref:12} Cannizzo J. K. , 1993, in: Accretion Disks in Compact
Stellar Systems, ed. J. Wheeler, World Scientific, Singapore, p. 6
\bibitem{ref:14} Hellier C., 2001, PASP 113, 469
\bibitem{ref:16} Ichikawa S., Osaki Y. 1992, PASJ 44, 15
\bibitem{ref:17} Hirose M., Osaki Y., 1990, PASJ 42, 135
\bibitem{ref:18} Kato T., Nogami D., Baba H. et al., 1996, PASJ, 48, L21
\bibitem{ref:19} Leibowitz E.M., Mendelson H., Bruch A. et al., 1994,
ApJ 421, 771
\bibitem{ref:20} Lin D.N.C., Papaloizou J.C.B, 1979, MNRAS 186, 799
\bibitem{ref:22} Lubow S.H., 1991, ApJ, 381, 259 
\bibitem{ref:24} Matsumoto K., Nogami D., Kato T. et al., 1998,
PASJ 50, 405
\bibitem{ref:26} Meyer F., Meyer-Hofmeister E., 1981, A\&A 104, L10
\bibitem{ref:27} Meyer F., Meyer-Hofmeister E., 1999, A\&A 341, L23
\bibitem{ref:28} Osaki Y., 1974, PASJ 26, 429
\bibitem{ref:30} Osaki Y., 1989, PASJ 41, 1005
\bibitem{ref:31} Osaki Y., 1995, PASJ 47, 47
\bibitem{ref:32} Osaki Y., 1996, PASP 108, 39
\bibitem{ref:33} Osaki Y., Meyer F., Meyer-Hofmeister E., 2001, A\&A
370, 488
\bibitem{ref:34} Paczynski B. 1977, ApJ 216, 822
\bibitem{ref:36} Papaloizou J.C.B., Pringle J.E., 1977, MNRAS 181, 441 
\bibitem{ref:38} Patterson J., McGraw J.T., Coleman et al. 1981, ApJ, 
248, 
1067
\bibitem{ref:40} Patterson J., Augusteijn T., Harvey et al., 1996,
PASP, 108, 748
\bibitem{ref:41} Smak J., 1993, Acta Astron. 43, 101
\bibitem{ref:42} Warner B., 1995, Cataclysmic Variable Stars,
Cambridge University Press 
\bibitem{ref:44} Whitehurst R., 1988, MNRAS, 232, 35
\bibitem{ref:46} Whitehurst R., 1994, MNRAS 266, 35
\end{thebibliography}
\end{document}